\documentclass{amia}
\usepackage{lipsum} %Remove if not needed
\usepackage{footnote}
\usepackage[colorlinks,allcolors=blue]{hyperref}

\begin{document}

\title{Evaluation of accessibility of open-source EHRs for visually impaired users}

\author{Megha M Moncy, BDS$^1$, Manya Pilli, BTech$^1$, Manasi Somasundaram, PharmD$^1$, Saptarshi Purkayastha, PhD$^1$, Cathy R. Fulton, DNP, RN, ANP-BC, FNP, BC$^1$ \newline}

\institutes{
    $^1$ Dept. of BioHealth Informatics, Indiana University Purdue University Indianapolis, Indianapolis, IN.
}

\maketitle

\section*{Abstract}

\textit{This study investigates the accessibility of open-source electronic health record (EHR) systems for individuals who are visually impaired or blind. Ensuring the accessibility of EHRs to visually impaired users is critical for the diversity, equity, and inclusion of all users. The study used a combination of automated and manual accessibility testing with screen readers to evaluate the accessibility of three widely used open-source EHR systems. We used three popular screen readers - JAWS (Windows), NVDA (Windows), and Apple VoiceOver (OSX) to evaluate accessibility. The evaluation revealed that although each of the three EHR systems was partially accessible, there is room for improvement, particularly regarding keyboard navigation and screen reader compatibility. The study concludes with recommendations for making EHR systems more inclusive for all users and more accessible.}

\section{Introduction}
Open source Electronic Health Records (EHRs) have revolutionized healthcare delivery by providing efficient access and management of patient information, resulting in improved decision-making, reduced medical errors, and enhanced quality of care \cite{atasoy2019digitization}. Their increased popularity is attributed to their cost-effectiveness and community-driven development \cite{shaikh2022open}. Even though EHRs have become ubiquitous in healthcare delivery, there has been a lack of attention to ensuring these systems are accessible to all users, including visually impaired individuals \cite{tsai2020effects}. There are an estimated 8 million visually impaired people in the USA, and many providers who need access to EHRs might be denied access due to visual impairment. Many of them face challenges when using EHRs due to their inability to read text, distinguish colors, or navigate complex interfaces \cite{killeen2023population}. The American Disability Act requires that anyone with disabilities should have equal opportunity, but EHR systems may be denying this opportunity due to accessibility problems. This can compromise the quality of healthcare that visually impaired users might give or receive and potentially worsen healthcare outcomes \cite{lagu2022not}. Therefore, it is critical to ensure that EHR systems are designed with accessibility in mind so that all patients can benefit from the advancements in digital healthcare technology \cite{oswal2021ideating}. Despite the significant population of visually impaired individuals, many open-source EHR systems have not been developed with their usability in mind \cite{farhan2023prototyping}. To provide inclusive and accessible healthcare delivery, EHR developers must prioritize accessibility in their designs.

OpenEMR, OpenMRS, and OSHERA VistA are three open-source EHRs that are widely used in healthcare delivery \cite{purkayastha2019comparison}. OpenEMR is primarily used in small to medium-sized practices due to its cost-effectiveness and community-driven development \cite{wang2022wearable}. OpenMRS, on the other hand, is designed to manage patient information in resource-limited settings and is widely used in developing countries \cite{mamlina2021openmrs}. Although the VA is replacing it, OSHERA VistA, which was initially developed by the US Department of Veterans Affairs (VA), still has a robust user base in the US and outside. Moreover, it is widely recognized as some of the most user-friendly and usable EHR systems in the US \cite{reisman2017ehrs}. There have been several studies examining the features and capabilities of open-source EHRs. However, a significant research gap exists in evaluating their accessibility for visually impaired users.

This study aims to evaluate the accessibility of three open-source EHR systems, namely OpenEMR, OpenMRS, and OSHERA VistA, for blind and low-vision users through the use of screen readers. JAWS, NVDA, and Apple VoiceOver are examples of popular screen readers used by visually impaired individuals to access digital content, including EHRs. In our research, we utilized these screen readers to evaluate the accessibility of OpenEMR, OpenMRS, and OSHERA VistA for blind and low-vision users. These screen readers provide audio feedback, allowing users to navigate interfaces and access content through keyboard commands \cite{kuber2020determining}. By using these screen readers to assess the accessibility of the EHR systems, we identified specific challenges faced by visually impaired users and provided recommendations for improving accessibility in EHR design. Our research findings can guide the creation of EHR systems that are more inclusive and accessible for visually impaired patients, ultimately leading to better healthcare delivery.

\section{Literature Review}
Open-source Electronic Health Record (EHR) systems have been widely adopted by healthcare systems globally because they offer cost-effective solutions for managing patient data, which are crucial components of modern healthcare systems. However, despite their increasing popularity, making these systems accessible to visually impaired users remains a significant challenge. Back in 2014, the ONC Health IT curriculum had clear learning outcomes to train informaticians to assess EHR systems with accessibility in mind \cite{mohan2014design}. However, the lack of published research and the lack of robust, large-scale accessibility evaluation of EHR systems makes us question the scale-up and reach of the curricular guidance.

OpenEMR is designed with accessibility in mind, but several studies have identified some accessibility issues that must be addressed. A study by Dimauro et al. (2018) also reported that OpenEMR had usability issues. The study identified problems with the system's keyboard navigation and labeling of controls \cite{dimauro2019personal}.

Although OpenMRS was not originally designed for visually impaired users, recent studies have shown that the system has significantly enhanced accessibility. For instance, a study by Verma et al. (2021) found that OpenMRS had good accessibility features, including keyboard navigation, focus management, labeling of controls, and color contrast \cite{verma2021openmrs}. Other studies highlight the potential of OpenMRS having good accessibility for visually impaired users \cite{schaadhardt2021understanding}. However, our findings with screen readers provide a different evaluation.

However, using EHRs has also introduced accessibility issues for individuals with disabilities, including visual, auditory, physical, and cognitive impairments. 
\begin{itemize}
    \item \underline{Visual Accessibility:}  One of the primary accessibility issues with EHRs is related to visual impairments. EHRs are typically designed with small fonts and low-contrast colors, making it difficult for individuals with visual impairments to read the content. Additionally, EHRs often require users to navigate through multiple screens and menus, which can be challenging for individuals with low vision or blindness.
    \item \underline{Auditory Accessibility:}  EHRs also pose accessibility issues for individuals with hearing impairments. Many EHRs rely on auditory alerts and notifications, which can be missed by individuals with hearing impairments.
    \item \underline{Physical Accessibility:} EHRs can also pose physical accessibility issues for individuals with mobility impairments. Many EHRs require users to navigate through multiple screens and menus, which can be challenging for individuals with limited dexterity or range of motion. Additionally, some EHRs require users to use a mouse or trackpad, which can be difficult for individuals with physical disabilities to operate.
    \item \underline{Cognitive Accessibility:} EHRs can also pose cognitive accessibility issues for individuals with cognitive impairments, such as learning disabilities or dementia. Many EHRs use complex medical terminology and require users to navigate through multiple screens and menus, which can be challenging for individuals with cognitive impairments to understand.
\end{itemize}

The accessibility of websites for visually impaired users has been extensively studied, and the findings can be useful in improving the accessibility of open-source EHRs \cite{beier2021increased}. Various methods have been identified to enhance the readability of web pages for people with low vision, such as increased letter spacing and wider letters. Moreover, better layout and design, consistent font usage, and color contrast are also considered important factors in improving accessibility \cite{luy2021toolkit}. Assistive technologies, such as screen readers that convert text into synthesized speech have been found to improve the usability of websites for the blind significantly. While there are multiple approaches to improving accessibility, screen readers are considered one of the easiest to implement. They offer a relatively simple and cost-effective way to enhance the usability of websites for visually impaired users. By converting text into speech, screen readers allow users to access and interact with web content without relying on visual cues \cite{elmqvist2023visualization}. Thus, ensuring EHR user interfaces work with screen readers can significantly improve accessibility and enable visually impaired individuals to benefit from these systems.

Screen readers are essential tools for visually impaired individuals to access and interact with digital content. Popular screen readers include JAWS, NVDA, and Apple VoiceOver. JAWS, a proprietary software developed by Freedom Scientific, is widely used in Windows-based applications due to its compatibility and advanced features \cite{mulyati2023development}. NVDA, an open-source screen reader, has gained popularity due to its cost-free accessibility and compatibility with Windows-based applications \cite{saha2023tutoria11y}. Apple VoiceOver, an integrated screen reader in Mac OS X, iOS, and iPad OS, provides seamless accessibility for Apple device users. These screen readers enable visually impaired users to access and interact with EHR systems, ensuring that the systems are inclusive and accessible to all users \cite{kumari2023employment}.

\section{Methodology}
\subsection*{Data Collection and Source of Information}
We ensured that we covered all the relevant literature on the accessibility of EHR systems by conducting a robust literature review. We utilized PubMed and IEEE Xplore databases to gather information from diverse sources. Furthermore, we examined the official websites of three open-source EHR systems to collect more details on their accessibility considerations and found little to no evidence that accessibility was actively considered by the developer community of these EHR systems. By incorporating these multiple sources of information, we aimed to provide a thorough and reliable analysis of screen reader accessibility in various open-source EHR platforms.

\subsection*{Methods}
The research methodology used in this study was developed in three distinct phases. The first phase involved deploying the open-source EHR software on institutional servers, using the available documentation from their official websites. This allowed us to create a controlled testing environment and ensured the EHR systems were tested similarly. In the second phase, we established evaluation criteria by identifying eight parameters through a literature review and qualitative content analysis, similar to our prior work comparing open-source EHR systems \cite{purkayastha2019comparison}. These parameters were chosen to encompass common tasks that users might need to perform when using an EHR system, such as logging in, finding a patient, creating a new password, logging out, entering and modifying demographics, entering medication orders, scheduling an appointment, and ordering a lab. Using a comprehensive set of evaluation criteria, we aimed to evaluate the EHR systems' accessibility thoroughly. In the final phase, we utilized three different screen readers - JAWS, NVDA, and Apple VoiceOver - to evaluate using the specified set of criteria. These screen readers were chosen based on their popularity among visually impaired users and their availability across different platforms. To better understand the functionality of these screen readers, we consulted various resources, including IEEE Xplore and Google Scholar.

\subsection*{Comparison Criteria}
Each open-source EHR system was evaluated using established criteria (see Table \ref{table:comparison}, ensuring the evaluation was objective and consistent. We tested each system with multiple screen readers and noted accessibility issues. This approach allowed for identifying any differences in accessibility between the three EHR systems and screen readers for blind users.

The results of this study can inform developers and designers to make EHR systems more accessible and inclusive for all users, including those with visual impairments. By doing so, EHR systems can better serve their intended purpose of improving patient care and outcomes by ensuring that all users can easily and effectively use them, regardless of their abilities.

\section{Results}

\begin{table}[!ht]
\centering
\caption{A Comparative Table Evaluating Screen Reader Accessibility in Various Open-Source EHR Platforms}
\label{table:comparison}
\begin{tabular}{|l|l|cc|cc|cl|}
\hline
\textbf{Open Source EHRs} & \textbf{TASK CRITERIA}        & \multicolumn{2}{l|}{\textbf{JAWS}}                                           & \multicolumn{2}{l|}{\textbf{NVDA}}                                           & \multicolumn{2}{l|}{\textbf{AppleVoiceOver}}                   \\ \hline
                          &                               & \multicolumn{1}{l|}{\textbf{Success}} & \multicolumn{1}{l|}{\textbf{Issues}} & \multicolumn{1}{l|}{\textbf{Success}} & \multicolumn{1}{l|}{\textbf{Issues}} & \multicolumn{1}{l|}{\textbf{Success}} & \textbf{Issues}        \\ \hline
\textbf{OpenEMR}          & Login                         & \multicolumn{1}{c|}{Yes}              & *                                    & \multicolumn{1}{c|}{Yes}              & *                                    & \multicolumn{1}{c|}{Yes}              & \multicolumn{1}{c|}{*} \\ \hline
\textbf{}                 & Find a patient                & \multicolumn{1}{c|}{Yes}              & \multicolumn{1}{l|}{}                & \multicolumn{1}{c|}{Yes}              & \multicolumn{1}{l|}{}                & \multicolumn{1}{c|}{Yes}              &                        \\ \hline
\textbf{}                 & Create a new patient          & \multicolumn{1}{c|}{Yes}              & *                                    & \multicolumn{1}{c|}{Yes}              & *                                    & \multicolumn{1}{c|}{Yes}              & \multicolumn{1}{c|}{*} \\ \hline
\textbf{}                 & Logout                        & \multicolumn{1}{c|}{No}               & *                                    & \multicolumn{1}{c|}{No}               & *                                    & \multicolumn{1}{c|}{No}               & \multicolumn{1}{c|}{*} \\ \hline
\textbf{}                 & Enter and modify demographics & \multicolumn{1}{c|}{Yes}              & \multicolumn{1}{l|}{}                & \multicolumn{1}{c|}{Yes}              & \multicolumn{1}{l|}{}                & \multicolumn{1}{c|}{Yes}              &                        \\ \hline
\textbf{}                 & Enter in medication order     & \multicolumn{1}{c|}{Yes}              & *                                    & \multicolumn{1}{c|}{Yes}              & *                                    & \multicolumn{1}{c|}{Yes}              & \multicolumn{1}{c|}{*} \\ \hline
\textbf{}                 & Schedule an appointment       & \multicolumn{1}{c|}{Yes}              &                                      & \multicolumn{1}{c|}{Yes}              &                                      & \multicolumn{1}{c|}{Yes}              & \multicolumn{1}{c|}{}  \\ \hline
\textbf{}                 & Order a lab                   & \multicolumn{1}{c|}{Yes}              & *                                    & \multicolumn{1}{c|}{Yes}              & *                                    & \multicolumn{1}{c|}{Yes}              & \multicolumn{1}{c|}{*} \\ \hline
\textbf{OpenMRS}          & Login                         & \multicolumn{1}{c|}{Yes}              & \multicolumn{1}{l|}{}                & \multicolumn{1}{c|}{Yes}              & *                                    & \multicolumn{1}{c|}{Yes}              & \multicolumn{1}{c|}{*} \\ \hline
\textbf{}                 & Find a patient                & \multicolumn{1}{c|}{Yes}              & *                                    & \multicolumn{1}{c|}{Yes}              & *                                    & \multicolumn{1}{c|}{Yes}              & \multicolumn{1}{c|}{*} \\ \hline
\textbf{}                 & Create a new patient          & \multicolumn{1}{c|}{Yes}              & *                                    & \multicolumn{1}{c|}{Yes}              & *                                    & \multicolumn{1}{c|}{Yes}              & \multicolumn{1}{c|}{}  \\ \hline
\textbf{}                 & Logout                        & \multicolumn{1}{c|}{Yes}              & \multicolumn{1}{l|}{}                & \multicolumn{1}{c|}{Yes}              & \multicolumn{1}{l|}{}                & \multicolumn{1}{c|}{Yes}              &                        \\ \hline
\textbf{}                 & Enter and modify demographics & \multicolumn{1}{c|}{Yes}              & *                                    & \multicolumn{1}{c|}{Yes}              & *                                    & \multicolumn{1}{c|}{Yes}              & \multicolumn{1}{c|}{*} \\ \hline
\textbf{}                 & Enter in medication order     & \multicolumn{1}{c|}{Yes}              & \multicolumn{1}{l|}{}                & \multicolumn{1}{c|}{Yes}              & \multicolumn{1}{l|}{}                & \multicolumn{1}{c|}{Yes}              &                        \\ \hline
\textbf{}                 & Schedule an appointment       & \multicolumn{1}{c|}{Yes}              & *                                    & \multicolumn{1}{c|}{Yes}              & *                                    & \multicolumn{1}{c|}{Yes}              & \multicolumn{1}{c|}{*} \\ \hline
\textbf{}                 & Order a lab                   & \multicolumn{1}{c|}{Yes}              & \multicolumn{1}{l|}{}                & \multicolumn{1}{c|}{Yes}              & \multicolumn{1}{l|}{}                & \multicolumn{1}{c|}{Yes}              &                        \\ \hline
\textbf{OSHERA VistA}     & Login                         & \multicolumn{1}{c|}{Yes}              & \multicolumn{1}{l|}{}                & \multicolumn{1}{c|}{Yes}              & \multicolumn{1}{l|}{}                & \multicolumn{1}{l|}{}                 &                        \\ \hline
                          & Find a patient                & \multicolumn{1}{c|}{Yes}              & *                                    & \multicolumn{1}{c|}{Yes}              & *                                    & \multicolumn{1}{l|}{}                 &                        \\ \hline
                          & Create a new patient          & \multicolumn{1}{c|}{No}               & *                                    & \multicolumn{1}{c|}{No}               & *                                    & \multicolumn{1}{l|}{}                 &                        \\ \hline
                          & Logout                        & \multicolumn{1}{c|}{No}               & *                                    & \multicolumn{1}{c|}{No}               & *                                    & \multicolumn{1}{l|}{}                 &                        \\ \hline
                          & Enter and modify demographics & \multicolumn{1}{c|}{No}               & *                                    & \multicolumn{1}{c|}{No}               & *                                    & \multicolumn{1}{l|}{}                 &                        \\ \hline
                          & Enter in medication order     & \multicolumn{1}{c|}{Yes}              & *                                    & \multicolumn{1}{c|}{Yes}              & *                                    & \multicolumn{1}{l|}{}                 &                        \\ \hline
                          & Schedule an appointment       & \multicolumn{1}{c|}{Yes}              & *                                    & \multicolumn{1}{c|}{Yes}              & *                                    & \multicolumn{1}{l|}{}                 &                        \\ \hline
                          & Order a lab                   & \multicolumn{1}{c|}{Yes}              & *                                    & \multicolumn{1}{c|}{Yes}              & *                                    & \multicolumn{1}{l|}{}                 &                        \\ \hline
                          
\end{tabular}\\
\footnotesize{* - issue details are provided in the sections below}\\
\end{table}

As seen in Table \ref{table:comparison}, all 3 EHR systems have issues with one or more screen readers. Although these EHR systems may perform well in terms of their intended functionality, there are still some accessibility issues that may hinder visually impaired users from utilizing them fully. It is important to note that the experience of using these systems may vary depending on the assistive technology being used, as different screen readers have varying features and functionalities.

We have identified Login, Find a patient, Create a new patient, Logout, Enter and modify demographics, Enter in medication order, and Schedule an appointment as critical parameters to evaluate the compatibility of EHRs with screen readers. These parameters are commonly used by healthcare professionals, and their compatibility with screen readers is crucial to ensure that visually impaired professionals can use EHRs effectively and efficiently. For instance, accurate reading of demographic fields and medication order fields is essential for healthcare professionals with visual impairments to enter data or modify patient records. Similarly, accurate reading of the schedule an appointment field is essential to enable visually impaired healthcare professionals to schedule appointments with patients. Therefore, evaluating EHRs compatibility with screen readers while performing these tasks is critical to ensure that visually impaired healthcare professionals can access and use EHRs effectively.

\subsection*{OpenEMR}
\begin{figure}[H]
  \centering
    \includegraphics[scale=0.25]{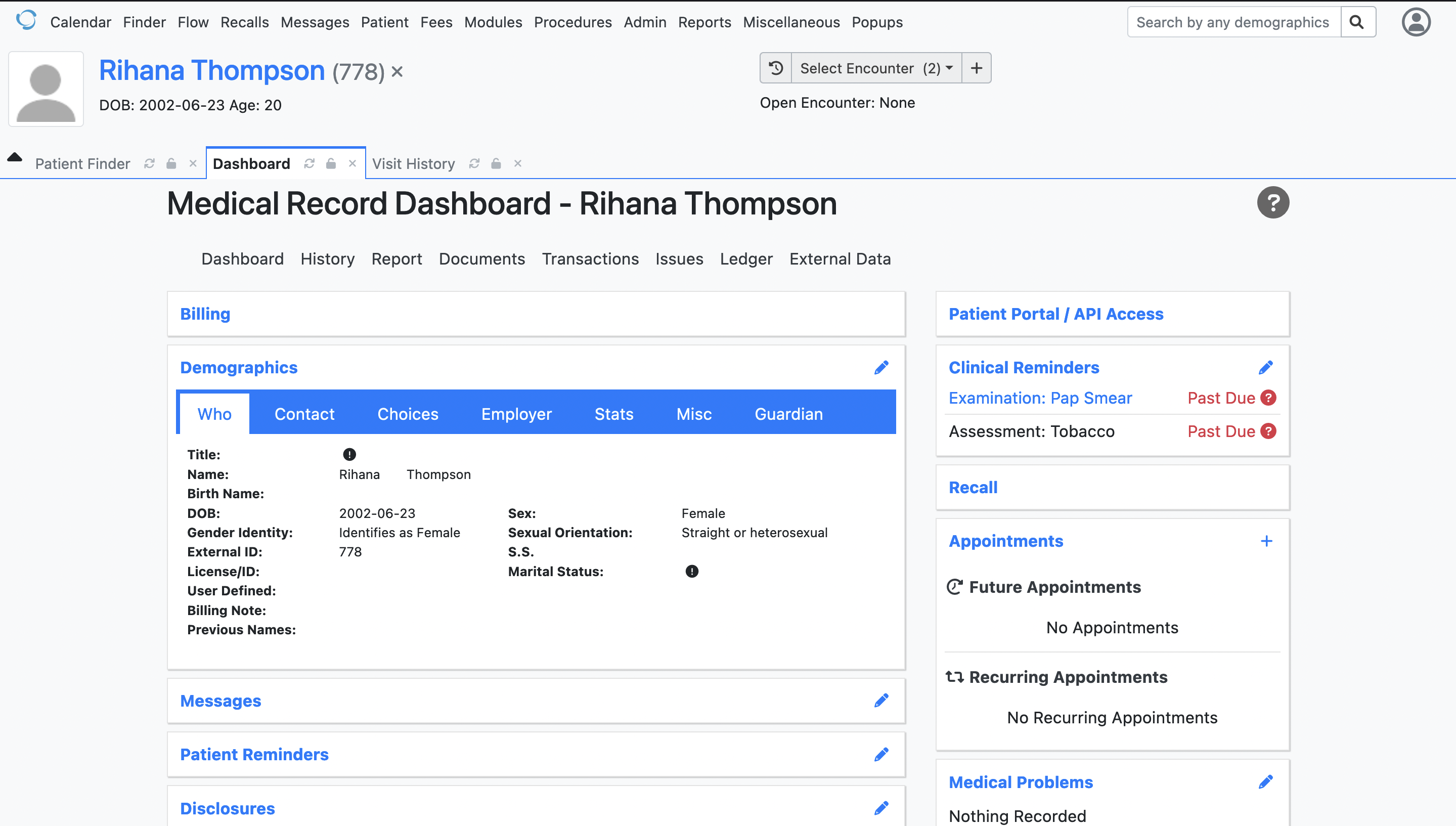}
 \caption{Dashboard of OpenEMR}
 \label{fig:OpenEMR}
\end{figure}
Based on our considered parameters, OpenEMR appears to be mostly compatible with popular screen readers such as JAWS, NVDA, and Apple Voiceover.

\begin{itemize}
    \item \underline{Login}: The screen readers read out the selected fields for login, finding a patient, and creating a new patient, which indicates that the EHR's basic accessibility functions are in place. Although screen readers read the login information accurately, they could not report error messages or provide feedback to users when incorrect information was entered in the login fields. 
    \item \underline{Create a new patient}: While creating a patient in OpenEMR, the consecutive tabs such as demographics, contact information, and so on were reliably reported by the three screen readers we tested (JAWS, NVDA, and Apple VoiceOver). This allowed users to navigate through different fields and enter the appropriate information easily. An exception is that the screen readers could not report which required field was left blank, making it challenging for visually impaired users to complete the patient creation process.
 
     \item \underline{Logout }: The process of logging out was found to be inefficient when using keyboard clicks, and it was discovered that users couldn't log out using the Tab key. Additionally, no audible confirmation was provided to indicate that the user had successfully logged out when using the cursor.
     
     \item \underline{Enter and modify demographics}: It's noteworthy that the use of screen readers proved to be efficient in assisting users with editing and updating patient demographics. This capability is particularly valuable when there is a need to include any missing information for a patient.

     \item \underline{Enter in medication order}: Entering medication details was a straightforward process. However, the medication list was not read out, which could create difficulties for visually impaired users to input medication information accurately.

     \item \underline{Order a lab}: During the lab ordering process, it was discovered that certain required options could not be accessed using the Tab key and required using a mouse. This presents a significant challenge for visually impaired users who rely on keyboard navigation to complete tasks. 

\end{itemize}

 While OpenEMR has some accessibility concerns, it appears to be a generally functional and efficient system for users who rely on screen readers. However, there may be room for further improvements to ensure all users can fully utilize the software's features.

\subsection*{OpenMRS}

\begin{figure}[H]
  \centering
    \includegraphics[scale=0.35]{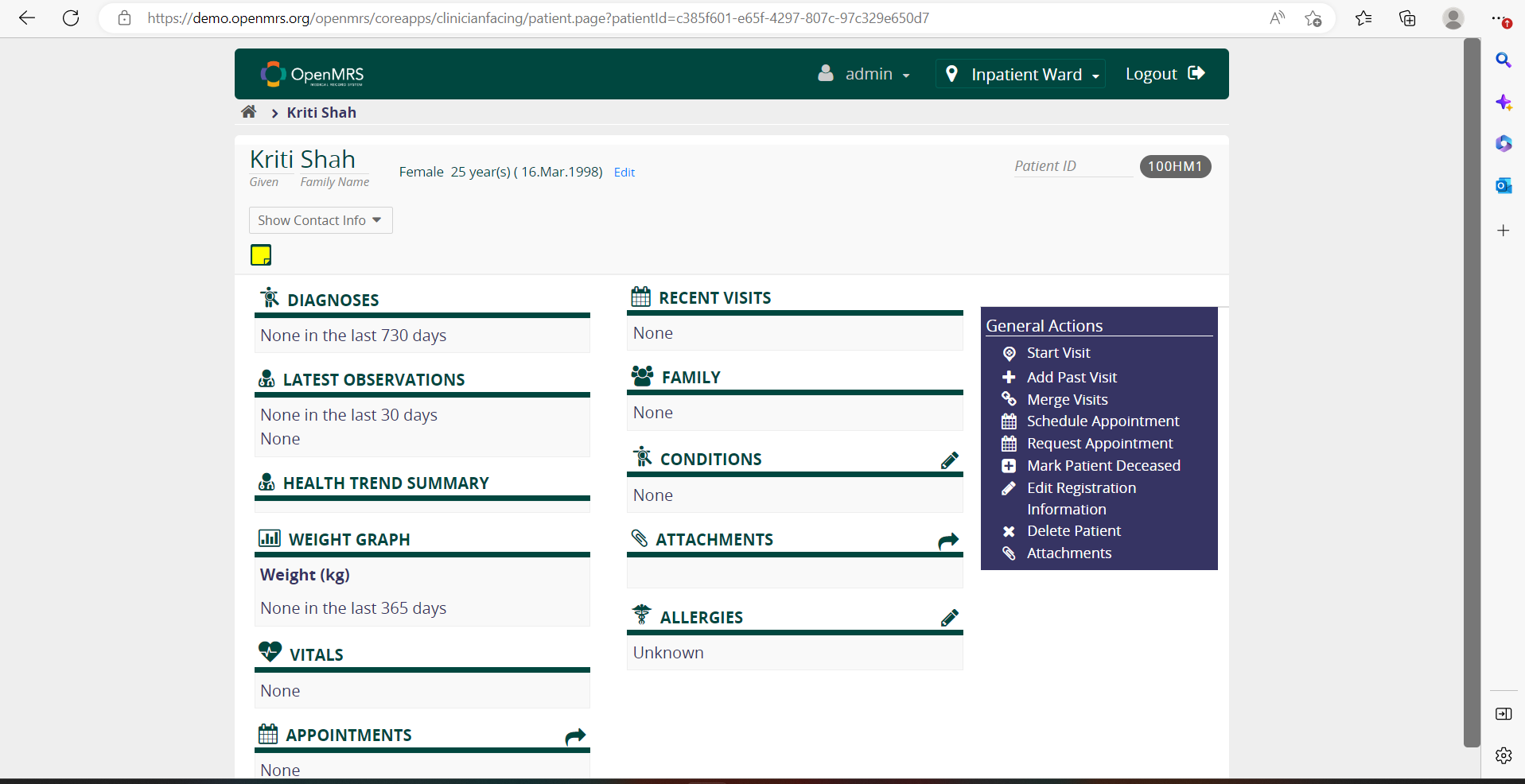}
 \caption{Dashboard of OpenMRS}
 \label{fig:OpenMRS}
\end{figure}

OpenMRS is an accessible platform that meets all the parameters considered. 
   \begin{itemize} 
   \item \underline{Login}: The login and log-out features were particularly efficient and user-friendly. However, we did observe a difference in performance among the three screen readers we tested (JAWS, NVDA, and Apple Voiceover) regarding the reporting of incorrect login information. While JAWS reported when incorrect information was entered into the system, NVDA and Apple Voiceover did not. NVDA does not allow users to navigate the department options and automatically selects the registration desk.
   \item \underline{Find a patient}: We found it easy to navigate and find a patient in OpenMRS, although the feature to select from a list of existing names is not compatible with the screen readers, making it difficult for users to find patients efficiently.
   \item \underline{Create a new patient}: Creating a new patient is relatively easy, although relationships cannot be entered using the tab key, which can slow down the data entry process.
    \item \underline{Enter and modify demographics}: Entering demographics in OpenMRS was relatively easy, as the screen readers provided clear and concise navigation options. When modifying demographics, the screen readers provide clear and concise navigation options, but entering conditions and allergies requires a mouse, which can be challenging for keyboard-only users. Also, the "Save form" and "Exit form" options are inaccessible on Apple Voiceover, which could confuse users who want to save or exit a form.
    \item \underline{Schedule an appointment}: Scheduling appointments works well with all screen readers, but entering the date and time using the keyboard is impossible, which could be a significant barrier for some users.
\end{itemize}
Overall, OpenMRS is an accessible platform that could benefit from further improvements, especially regarding keyboard navigation and screen reader compatibility for certain features.

\subsection*{OSHERA VistA}

\begin{figure}[H]
  \centering
    \includegraphics[scale=0.5]{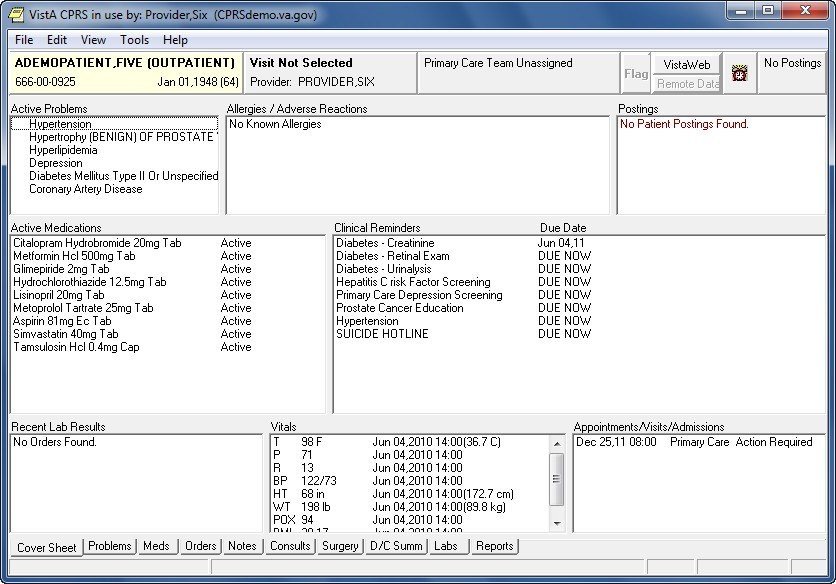}
 \caption{Dashboard of VistA}
 \label{fig:VistA}
\end{figure}

OSHERA VistA has several accessibility features that work well, but there are also some areas where improvement is needed. We evaluated using JAWS and NVDA, as the system is incompatible with Apple VoiceOver. 

\begin{itemize}
    \item \underline{Login}: The login feature in OSHERA VistA worked efficiently with both JAWS and NVDA, without any significant issues. 

    \item\underline{Find a patient}: When finding a patient, we found that NVDA performed better than JAWS as the latter could not read out the names being scrolled through in the drop-down list. 

    \item\underline{Create a new patient}: We faced an issue creating a new patient with both JAWS and NVDA on OSHERA VistA. These screen readers did not read out the respective feature despite accessing all the tabs on the interface. 

    \item\underline{Logout}: There was an issue with the logout feature in the EHR system. Specifically, there was no option to log out and sign in with a different account, and the system directly exited from the entire EHR.

    \item\underline{Enter and modify demographics}: Although it was possible to review the selected patient's demographics, there is no option to enter new information or modify the demographics that are reported by both the screen readers (JAWS and NVDA).

    \item\underline{Enter in medication order}: In the medication order feature, JAWS incorrectly guided the user to use the Ctrl + Page Up keys to shift pages, and neither JAWS nor NVDA read out the missing required fields when the order failed to save. These issues can make it difficult for users to complete medication orders accurately and efficiently.

    \item\underline{Schedule an appointment}: Scheduling an appointment had pertinent accessibility issues, such as the inability to move past the "Reason for Request" box with the tab key and the lack of feedback from both screen readers upon quitting. Also, JAWS and NVDA failed to read out the number being selected in the calendar when picking a date, which could make it difficult for users to select the correct date for their appointment.

    \item\underline{Order a lab}: The screen readers provided clear and concise navigation options, making it easy for users to fill in the appropriate fields. However, both screen readers failed to read the message of an unsaved order dialogue box when quitting a window. Finally, the lack of feedback from both screen readers upon successfully adding a lab test was another pertinent issue, as users may not know whether the test has been added successfully, which could lead to confusion.

\end{itemize}

\section{Discussion}
Much effort has been spent making internet websites accessible to the blind or visually challenged \cite{kusumaningayu2017web}. While various tools are available to improve accessibility, including screen readers, Braille displays, software for magnifying objects, Windows Voice Recognition, Microsoft Narrator, and Dragon Naturally Speaking, not all individuals with visual impairments are proficient in using computers \cite{gaggi2019accessibility}. To enhance the accessibility of websites for those with visual impairments, it is necessary to modify the color contrast and font sizes, use informative headers, and ensure that the content is straightforward to access with a keyboard \cite{ara2023integrated}. However, choosing the right tools to improve accessibility is crucial. Among the various tools available, we have chosen JAWS, NVDA, and Apple VoiceOver as they are widely used and have proven to be effective in enhancing the accessibility of websites. They are effective, compatible, and popular among blind users. By using these tools, we can enhance the accessibility of websites and other digital content, making it easier for individuals with visual impairments to access the same information as those without such impairments.

The small user base of blind healthcare providers may contribute to why vendors are not prioritizing accessibility improvements for this particular user group. With a small user base, vendors may not see the potential return on investment for developing more accessible tools and features. However, this mindset overlooks the significant impact that such enhancements could have on the usability and productivity of blind providers who do exist. Therefore, while developing and implementing EHR systems, accessibility is crucial. A wide variety of features are available in JAWS and NVDA, including sophisticated navigation, customizability, and compatibility for several file formats and web browsers. Although it supports certain accessibility functions, such as basic navigation and typing, Apple VoiceOver was found to be not as advanced as JAWS and NVDA.

We have noticed that OpenMRS and OpenEMR both provide high levels of accessibility for individuals who are blind or visually impaired. Both EHRs have user interfaces that enable visually impaired users to access the system's content and operate the interface, such as text-to-speech and keyboard navigation. The analysis demonstrates that designing an EHR system with accessibility for users who are blind or visually impaired is essential. They have all the required key components that blind users require, which can make it easier for them to access the system's information.

When it comes to procedures like logging into the account, ordering a lab and medications, scheduling an appointment for a patient, editing their demographics and contact details, taking vital signs, etc., they were all easily completed consistently. By providing an audible description of the user interface and allowing users to navigate through menus, links, and other elements, screen readers increase the usability of open-source EHRs for people with visual impairments, ensuring they can interact with the system's user interface and input data accurately. Additionally, screen readers can help identify compatibility issues that may prevent users with visual impairments from using the system.

The study found a few accessibility barriers, including the complex interfaces and screen readers' sporadic partial completion of recitations, which made them difficult for users to comprehend and perceive. The most accessible EHR was determined to be OpenMRS, whereas OpenEMR and OSHERA VistA had some accessibility issues regarding compatibility with Apple VoiceOver, which was later corrected and evaluated as a part of the study. EHRs frequently have intricate user interfaces that may not be designed with screen readers in mind, making them challenging to navigate and interact with. This was evident in OpenMRS, where we were unable to use the keyboard to mention a patient's relationship and were unable to exit the form while editing the patient's demographics and contact information. For example, in OpenEMR, the screen reader could not recite that the login password was incorrect for the user to understand that they were typing the wrong password. This is because open-source EHRs are developed by a diverse range of contributors, so there may be inconsistencies in the way that different components of the software are labeled and organized. The patient finding was challenging in OpenMRS since we needed a mouse to scroll down. However, only a limited number of tasks in each of the three EHRs required the usage of a mouse rather than the keyboard's tab key.

Despite the valuable insights derived from this study, a key limitation lies in our inability to capture the perspectives of visually impaired individuals directly. This introduces potential bias in our understanding of screen reader use with EHRs, as our findings may not entirely align with the actual user experiences. Future research endeavors should seek to address this gap by actively incorporating the experiences and feedback of visually impaired individuals using EHRs with screen readers. Incorporating their feedback will allow us to more accurately and empathetically design, test, and improve EHR systems for those with visual impairments.

\section{Conclusion}
In conclusion, this study highlights the importance of accessibility in open-source EHR systems for users who are blind or visually impaired. The findings suggest that developers can make EHR systems more inclusive and accessible by incorporating text-to-speech and keyboard navigation features. Additionally, users and organizations should prioritize accessibility when selecting and implementing EHR systems to ensure all users can access and benefit from the system's content. Our next steps are to conduct wider functionality testing across these EHRs and include proprietary EHRs like Epic, Meditech, eClinicalWorks, and Cerner.

The study shows that all three open-source EHR systems evaluated offer a respectable level of accessibility for visually impaired users. However, there is still room for improvement in specific areas, such as the user interface of OSHERA VistA and the forms and navigation of OpenEMR. Comparatively, OpenMRS was found to have the fewest accessibility concerns among the three systems.
Overall, this study provides valuable insights into how open-source EHR systems can be made more accessible for individuals with visual impairments. It emphasizes the need to prioritize accessibility in EHR system design and implementation to improve access to healthcare information and services for individuals with disabilities. The findings presented in this research paper can serve as a valuable resource for developers, users, and organizations looking to design and implement accessible EHR systems that prioritize the needs of visually impaired users.

\section*{Acknowledgments}
We acknowledge the invaluable contributions of our teaching assistants, tech staff, and instructors supporting the deployment of the Educational EHR for which we started the accessibility review. We acknowledge Pallavi Singh, who helped review prior work on EHR accessibility. We acknowledge the support of the IUPUI STEM Education Innovation \& Research Institute's 2022 seed grant.

% References as numbers
\makeatletter
\renewcommand{\@biblabel}[1]{\hfill #1.}
\makeatother

% unstr is used to keep citation order
\bibliographystyle{vancouver}
\bibliography{amia}  

\end{document}